\documentclass[conference]{IEEEtran}

\ifCLASSINFOpdf
   \usepackage[pdftex]{graphicx}
\else
   \usepackage[dvips]{graphicx}
\fi

\ifCLASSOPTIONcomsoc
  \usepackage[caption=false,font=normalsize,labelfont=sf,textfont=sf]{subfig}
\else
  \usepackage[caption=false,font=footnotesize]{subfig}
\fi

\usepackage{footnote}
\usepackage{nopageno}
\usepackage{float}
\usepackage{url}
\usepackage{amssymb,amsfonts}
\usepackage{amsmath}

\usepackage{graphicx}
\usepackage{mathptmx}  
\usepackage{epstopdf}
\usepackage{textcomp}
\usepackage{xcolor}
\usepackage{xspace}
\usepackage[absolute,showboxes]{textpos}

\usepackage{multirow}
\usepackage{multicol}
\usepackage{array} 
\usepackage{cite}
\usepackage{algorithm,algorithmic}
\usepackage{amsmath}
\usepackage[cmintegrals]{newtxmath}

\usepackage[utf8]{inputenc}
\usepackage{blindtext}
\usepackage{csquotes}

\def\BibTeX{{\rm B\kern-.05em{\sc i\kern-.025em b}\kern-.08em
    T\kern-.1667em\lower.7ex\hbox{E}\kern-.125emX}}

\begin{document}
\pagenumbering{gobble}

\title{\huge Learning the APT Kill Chain: Temporal Reasoning over Provenance Data for Attack Stage Estimation}

\author{

Trung V. Phan and Thomas Bauschert \\
\IEEEauthorblockA{Chair of Communication Networks, Technische Universit{\"a}t Chemnitz,  09126 Chemnitz, Germany}
Email: trung.phan-van@etit.tu-chemnitz.de, thomas.bauschert@etit.tu-chemnitz.de

}

% make the title area
\maketitle

\begin{abstract}
Advanced Persistent Threats (APTs) evolve through multiple stages, each exhibiting distinct temporal and structural behaviors. Accurate stage estimation is critical for enabling adaptive cyber defense. This paper presents \textit{StageFinder}, a temporal-graph learning framework for multi-stage attack progression inference from fused host and network provenance data. Provenance graphs are encoded using a graph neural network to capture structural dependencies among processes, files, and connections, while a long short-term memory (LSTM) model learns temporal dynamics to estimate stage probabilities aligned with the MITRE ATT\&CK framework. The model is pretrained on the DARPA OpTC dataset and fine-tuned on labeled DARPA Transparent Computing data. Experimental results demonstrate that StageFinder achieves a macro F1-score of 0.96 and reduces prediction volatility by 31\% compared to state-of-the-art baselines (Cyberian, NetGuardian). These results highlight the effectiveness of fused provenance–temporal learning for accurate and stable APT stage inference.
\end{abstract}

\IEEEpeerreviewmaketitle

\pagestyle{headings}
\setcounter{page}{1}
\pagenumbering{arabic}

\section{Introduction}\label{Introduction}
\begin{figure}
\centering
\includegraphics[width=0.45\textwidth]{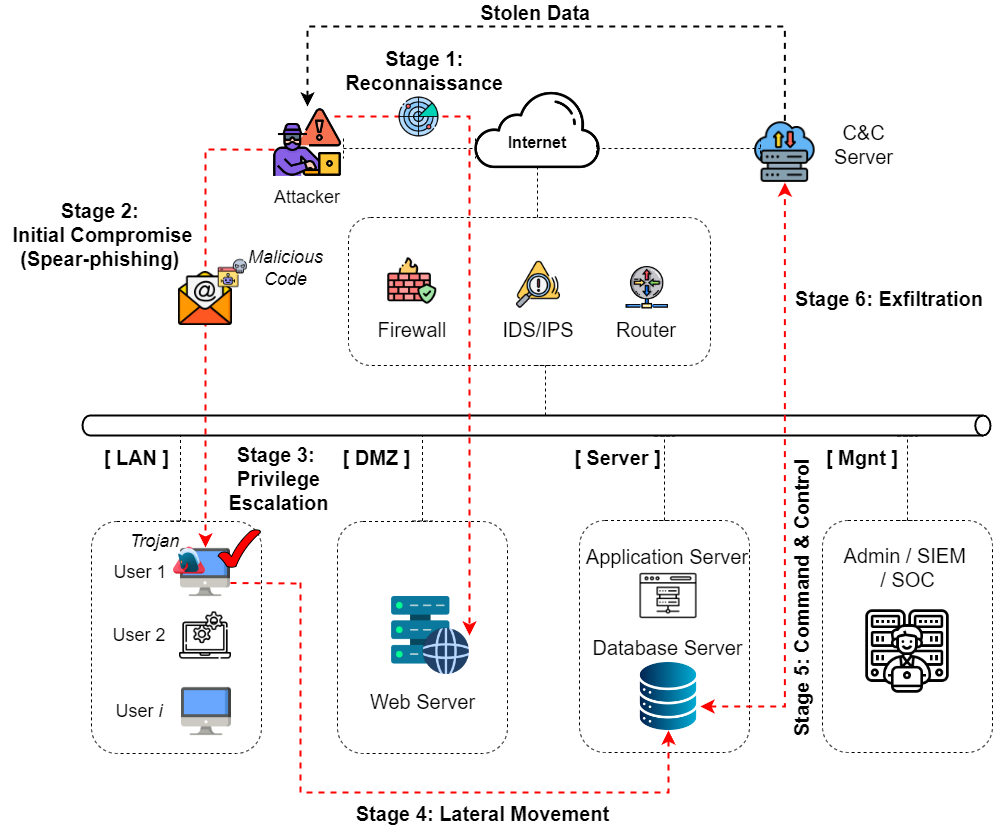}
\caption{An example of an APT attack towards an enterprise network.}
\label{fig:APT_Attack_to_Enterprise}
\end{figure}

Advanced Persistent Threats (APTs)~\cite{APTDetectionSurvey1} have emerged as one of the most formidable challenges in modern cybersecurity, targeting enterprise, governmental, and critical infrastructure networks. Unlike opportunistic or commodity malware, APTs are characterized by stealth, long dwell times, and multi-stage progressions designed to achieve long-term objectives such as data exfiltration, espionage, or operational disruption. As illustrated in Figure~\ref{fig:APT_Attack_to_Enterprise} (inspired by the MITRE ATT\&CK Enterprise matrix~\cite{mitre_attack}), a typical APT campaign begins with \textit{Reconnaissance}, followed by \textit{Initial Compromise}, \textit{Privilege Escalation}, \textit{Lateral Movement}, \textit{Command \& Control}, and ultimately \textit{Exfiltration}. Each stage leaves only subtle traces, often interleaved with benign activities, which makes accurate detection and interpretation particularly challenging~\cite{APTDetectionSurvey1}. 

Traditional signature-based intrusion detection and prevention systems (IDS/IPS) are effective against known exploits but fail to recognize novel or evolving Tactics, Techniques, and Procedures (TTPs)~\cite{DeepLearningforAPTDetection}. Anomaly-based methods expand coverage to unseen threats but frequently suffer from high false positive rates and lack contextual awareness of multi-step attack progressions~\cite{IDSleveragingHostData}. Furthermore, the low-and-slow behavior of APTs distributes weak indicators sparsely across logs and hosts, making causal inference and event correlation difficult using conventional detection pipelines~\cite{ProvenanceIDS}.

Each APT stage manifests a distinct behavioral signature. During \textit{Reconnaissance}, adversaries perform network scanning, service enumeration, and probing to map potential entry points. The \textit{Initial Compromise} phase often employs spear-phishing or malicious document delivery to gain initial access, while \textit{Privilege Escalation} involves exploiting local vulnerabilities to obtain elevated privileges. \textit{Lateral Movement} relies on credential reuse and remote code execution to spread laterally, and finally, \textit{Exfiltration} involves covert data transfer or command-and-control (C2) communication to external servers. Detecting and classifying these evolving phases---a process known as \textit{APT stage estimation}---is crucial for enabling adaptive, context-aware defenses. A stage-aware system can, for instance, apply selective monitoring during reconnaissance and escalate to aggressive containment during lateral movement, thereby improving response precision and reducing both false alarms and missed detections~\cite{APTDetectionSurvey1}.

This paper proposes \textit{StageFinder}, a temporal-graph learning framework for multi-stage attack progression inference. StageFinder continuously collects host-level system logs and transforms them into \textit{provenance graphs} that capture causal and temporal dependencies among system entities such as processes, files, users, and sockets. To incorporate broader situational context, an early fusion mechanism integrates network-layer alerts---for example, those produced by IDS or firewall systems---directly into the provenance graph. Each alert is modeled as a first-class node linked to the relevant host entities, preserving semantic relationships between network anomalies and local activities. The resulting fused provenance graph expresses both intra-host and inter-network dependencies in a unified causal space.

A graph neural network (GNN) encoder extracts low-dimensional embeddings from these fused graphs to represent structural and contextual patterns. These embeddings are then processed by a long short-term memory (LSTM) model that captures temporal dynamics and estimates the attacker’s probabilistic stage in the kill chain. Evaluations conducted on the DARPA Transparent Computing (TC)~\cite{DARPA_TC_Dataset} and Operationally Transparent Cyber (OpTC)~\cite{darpa_optc_dataset} datasets demonstrate that StageFinder achieves a macro F1-score of 0.96 and reduces prediction volatility by 31\% compared with state-of-the-art APT stage classification methods such as \textit{Cyberian}~\cite{Cyberian} and \textit{NetGuardian}~\cite{NetGuardian}. These results confirm the effectiveness of combining provenance-based graph modeling and temporal reasoning for accurate, interpretable, and stable multi-stage intrusion understanding.

\section{Related Work}
Research on Advanced Persistent Threat (APT) detection increasingly focuses on identifying and classifying attack stages, which reflect the temporal and structural evolution of an intrusion. Early work treated stage inference as a probabilistic state-transition problem, employing Hidden Markov Models (HMMs) to infer hidden attacker states from sequences of system logs~\cite{MaPLOS2024}. Although the model captures basic temporal dependencies, it relies on handcrafted features and fixed transition probabilities, limiting its adaptability to new or polymorphic attack behaviors.

With the availability of richer host-level telemetry, deep temporal models have become dominant. For example, Cyberian~\cite{Cyberian} applied LSTMs to the DARPA~Transparent Computing (TC) dataset for multi-stage intrusion classification, outperforming rule-based systems in precision and recall. However, purely sequential models fail to encode causal relations among entities---such as processes, files, and sockets---that are essential for distinguishing between subtle transitions (e.g., lateral movement versus command-and-control). Meanwhile, NetGuardian~\cite{NetGuardian} addressed this issue through a stage-specific ensemble in which dedicated classifiers target individual phases of the attack lifecycle. This modular structure improves interpretability but requires manual feature engineering and cannot generalize to unseen attack chains. To enhance temporal reasoning, Do~Xuan \textit{et al.}~\cite{DoXuanSciRep2024} combined BiLSTMs with attention mechanisms to highlight critical subsequences in network traffic. Despite its improved accuracy, these methods still treat host and network logs as independent event streams, overlooking causal dependencies across entities and systems.

To capture such relationships, provenance- and graph-based models have gained traction. System provenance provides a structured abstraction of entity interactions through event-driven data flows~\cite{ProvenanceIDS}, offering the context needed to reconstruct attack paths and understand stage transitions. Bahar \textit{et al.}~\cite{BaharSTGNN2025} proposed a spatio–temporal graph neural network that learns topological and temporal correlations for APT detection, while Wang \textit{et al.}~\cite{WangDynProv2025} developed a dynamic graph embedding approach that updates node representations as attacks evolve. Complementary runtime frameworks, such as Prographer~\cite{Prographer}, leverage provenance graphs for anomaly detection through graph compression and incremental analysis, enabling scalable monitoring of long-running systems. Zimba \textit{et al.}~\cite{Zimba2020} extended this direction using semi-supervised complex network modeling to infer APT phase transitions under limited labeled data. These studies confirm that causal graph representations and temporal reasoning jointly improve detection fidelity and temporal stability in APT stage classification.

\textit{Summary and Motivation:} Existing APT stage inference techniques exhibit complementary strengths yet face persistent limitations. Sequential models (e.g., Cyberian) effectively capture temporal evolution but ignore structural causality; stage-specific frameworks (e.g., NetGuardian) improve interpretability but lack unified temporal coherence; and graph-based methods excel at structural reasoning but often overlook multi-modal temporal dynamics. \textit{StageFinder} addresses these challenges through a unified temporal-graph learning approach that fuses host and network provenance data. By combining GNN-based structural encoding with LSTM-based temporal modeling, StageFinder achieves accurate, context-aware, and temporally stable estimation of APT progression across the enterprise environment.

\begin{figure*}
\centering
\includegraphics[width=1.0\textwidth]{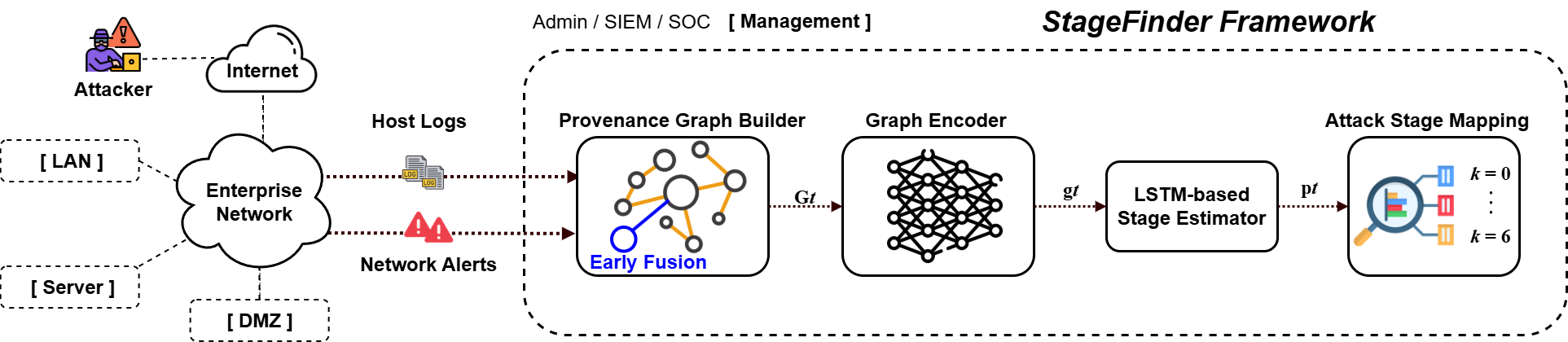}
\caption{Data and control flow of the \textit{StageFinder} framework. Host logs and network alerts are fused into a provenance graph, encoded by a GNN, and analyzed by an LSTM-based Stage Estimator, with the Attack Stage Mapping producing discrete APT stages.}
\label{fig:StageFinder}
\end{figure*}

\section{Design of the StageFinder Framework}\label{StageFinderFramework}
This section details the design of the proposed \textit{StageFinder} framework, which operates as a sequential processing pipeline within an enterprise network environment. As shown in Figure~\ref{fig:StageFinder}, the system ingests multi-source telemetry, constructs fused provenance graphs, encodes them via a GNN, and performs temporal stage estimation through an LSTM-based model, followed by interpretable attack stage mapping.

\subsection{Network Environment}
The pipeline begins with data collection in a controlled enterprise environment logically divided into four zones—local area network (LAN), demilitarized zone (DMZ), server zone, and management zone. The LAN contains employee workstations and internal services that often serve as entry points for phishing or drive-by compromise. The DMZ hosts externally accessible services (web, mail, VPN) and acts as a buffer between external and internal networks. The server zone stores critical business assets such as database and authentication servers, which are common APT targets during later stages. The management zone contains centralized monitoring and orchestration components, where the \textit{StageFinder} framework is deployed to collect telemetry from all other zones while remaining isolated from external access. Network traffic is filtered by perimeter firewalls and analyzed by IDS/IPS sensors (e.g., Zeek), ensuring that both host-level and network-level data are available for subsequent fusion.

\subsection{Early Fusion of Host and Network Provenance Data}\label{subsec:earlyfusion}
Next, StageFinder performs \textit{early fusion} to unify host and network data into provenance graphs. APTs typically manifest through coordinated host and network behaviors—where host telemetry (e.g., process creation, file I/O) alone cannot reveal external C2 or lateral movement, and network alerts lack causal attribution to specific processes. To bridge this gap, the fusion mechanism (highlighted in blue in Figure~\ref{fig:StageFinder}) integrates logs and alerts directly during provenance graph construction. Each process node (e.g., \textit{wget.exe}, \textit{powershell.exe}) is linked to IDS or firewall alert nodes when causal evidence of related activity is detected. Each alert node includes metadata (e.g., \textit{signature}, \textit{severity}) and is connected via time-ordered edges, preserving causality.  

This fine-grained integration provides three major benefits: (\textit{i}) \textit{Causal completeness}—each fused subgraph captures both intra-host dependencies (process–file–registry) and inter-host communications, enabling learning models to reason over full attack chains; (\textit{ii}) \textit{Contextual consistency}—fusion occurs prior to feature extraction, allowing GNN encoders to learn coherent representations across modalities; and (\textit{iii}) \textit{Enhanced stage awareness}—temporal alignment of host and network evidence improves the discrimination of key APT stages such as initial compromise, lateral movement, and data exfiltration. A concrete example of the proposed early-fusion mechanism is provided in the following section.

\begin{table}
\centering
\caption{Example captured host log events captured by Sysmon.}
\label{tab:sysmon_events}
\begin{tabular}{|>{\arraybackslash}p{1.8cm}|>{\arraybackslash}p{6.0cm}|}
\hline
 \textbf{Sysmon Event} & \textbf{Description} \\
\hline
 ProcessCreate & \texttt{powershell.exe} launches \texttt{wget.exe} \\
\hline
 FileCreate & \texttt{wget.exe} creates file \texttt{payload.exe} \\
\hline
 ProcessCreate & \texttt{payload.exe} starts execution \\
\hline
\end{tabular}
\end{table}

\subsection{Provenance Graph Builder}
The provenance graph builder then parses and correlates raw host logs and network alerts into a unified causal graph representation. For each time window \( t \), it constructs a fused provenance graph \( G_t = (V_t, E_t) \), where nodes \( V_t \) denote entities (processes, files, sockets, IP addresses, and alert events), and edges \( E_t \) represent causal or temporal dependencies such as \emph{read}, \emph{write}, \emph{spawn}, \emph{connect}, and \emph{triggered\_by}. Host and network data are integrated through the proposed \textit{early-fusion} strategy, which connects provenance and alert information directly during graph construction to preserve causal and temporal consistency.

\begin{figure}
\centering
\includegraphics[width=0.5\textwidth]{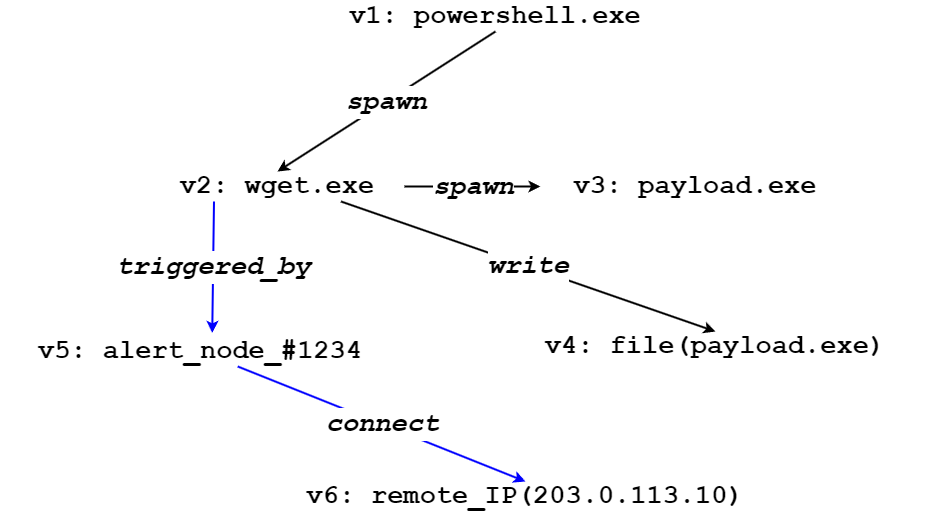}
\caption{An example of a provenance graph construction with the early fusion.}
\label{fig:Early_Fusion_Example}
\end{figure}

\textbf{Example:} Consider a scenario where a suspicious process on a workstation (\texttt{10.1.1.45}) downloads and executes a file from a remote server. As summarized in Table~\ref{tab:sysmon_events}, Sysmon logs record process creation and file operations, forming intra-host relations such as \texttt{powershell.exe} spawning \texttt{wget.exe}, which writes and executes \texttt{payload.exe}. Concurrently, a Zeek IDS generates an alert (\texttt{ET TROJAN Possible Malicious EXE Download}) between the local host and external IP \texttt{203.0.113.10}. The builder adds corresponding alert and IP nodes and links them to the responsible process via directed edges (\emph{triggered\_by}, \emph{connect}). The resulting fused subgraph thus unifies host-level causality with network evidence, enabling reasoning over complete attack chains rather than isolated events, as shown in Figure \ref{fig:Early_Fusion_Example}.

\subsection{Graph Encoder}
\label{sec:graph_encoder}
After graph construction, StageFinder transforms each fused graph into a low-dimensional embedding through the \textit{graph encoder}. The fused provenance graph is defined as 
$G_t = (V_t, E_t, X_t, Z_t)$, where $V_t$ and $E_t$ denote the sets of nodes (processes, files, sockets, users, alerts) and directed edges, respectively. 
Each node $v_i \!\in\! V_t$ and edge $e_{ij}\!\in\!E_t$ is associated with feature vectors $\mathbf{x}_i \!\in\! \mathbb{R}^{d_x}$ and $\mathbf{z}_{ij} \!\in\! \mathbb{R}^{d_e}$, forming 
$X_t = \{\mathbf{x}_i\}$ and $Z_t = \{\mathbf{z}_{ij}\}$. 
The goal is to encode $G_t$ into a low-dimensional graph embedding 
$g_t \!\in\! \mathbb{R}^{d_g}$.

\subsubsection{Node Feature Initialization}
Each node $v_i$ is initialized with a feature vector $\mathbf{x}_i \in \mathbb{R}^{d_x}$ encoding structural, semantic, and temporal attributes. 
We distinguish two primary node categories: host-log entities and network-alert entities.

\paragraph{Host Log Entities}
For system-derived nodes (process, file, socket, user, host), features are defined as
$
\mathbf{x}_i^{(\text{host})} = 
\big[\, \phi_{\text{type}} \;\Vert\; \phi_{\text{cmd}} \;\Vert\; \phi_{\text{user}} \;\Vert\; \phi_{\text{time}} \;\Vert\; \phi_{\text{stat}} \,\big],
\label{eq:host_feature_vector}
$
where $\phi_{\text{type}}$ is a one-hot entity-type encoding, 
$\phi_{\text{cmd}}$ is a TF--IDF embedding of command strings or file names, 
$\phi_{\text{user}}$ encodes user or privilege context, 
$\phi_{\text{time}}$ is a timestamp embedding, and 
$\phi_{\text{stat}}$ contains statistical indicators such as node degree or event counts. 

\paragraph{Network Alert Entities}
For alert nodes derived from IDS or firewall notifications, features are represented as
$
\mathbf{x}_i^{(\text{alert})} =
\big[\, \phi_{\text{sig}} \;\Vert\; \phi_{\text{sev}} \;\Vert\; 
\phi_{\text{proto}} \;\Vert\; \phi_{\text{net}} \;\Vert\; 
\phi_{\text{time}} \,\big],
\label{eq:alert_feature_vector}
$
where $\phi_{\text{sig}}$ encodes alert signature or textual message, 
$\phi_{\text{sev}}$ is a normalized severity score, 
$\phi_{\text{proto}}$ specifies protocol and flow direction, 
$\phi_{\text{net}}$ encodes IP/port or subnet context, 
and $\phi_{\text{time}}$ encodes relative time.

\subsubsection{Edge Feature Initialization}
Each directed edge $e_{ij} = (v_i, v_j, \tau_{ij}) \in E_t$ encodes a causal or communicative relation between two entities. Edges are represented by feature vectors $\mathbf{z}_{ij} \in \mathbb{R}^{d_e}$ to capture relation-specific semantics.

\paragraph{Host Log Edges}
Edges derived from system-level events are initialized as
$
\mathbf{z}_{ij}^{(\text{host})} =
\left[\, 
\psi_{\text{type}} \;\middle\|\;
\psi_{\text{freq}} \;\middle\|\;
\psi_{\text{size}} \;\middle\|\;
\psi_{\text{time}} 
\,\right],
$
where $\psi_{\text{type}}$ encodes the event type (read, write, exec, send, recv, etc.), 
$\psi_{\text{freq}}$ is the normalized interaction frequency, 
$\psi_{\text{size}}$ denotes log-normalized byte volume, 
and $\psi_{\text{time}}$ is the temporal offset within the window. 

\paragraph{Network Alert Edges}
Alert-related edges link alerts to affected hosts or sockets:
$
\mathbf{z}_{ij}^{(\text{alert})} =
\big[\, \psi_{\text{atype}} \;\Vert\; \psi_{\text{sev}} \;\Vert\; \psi_{\text{proto}} \;\Vert\; \psi_{\text{time}} \,\big],
$
where $\psi_{\text{atype}}$ encodes the alert category, 
$\psi_{\text{sev}}$ captures alert severity, 
$\psi_{\text{proto}}$ specifies protocol context, 
and $\psi_{\text{time}}$ encodes temporal recency. 

All continuous attributes are standardized using z-score normalization and projected into a shared latent space:
$
\tilde{\mathbf{x}}_i = W_x\mathbf{x}_i + b_x,
\qquad
\tilde{\mathbf{z}}_{ij} = W_z\mathbf{z}_{ij} + b_z,
\label{eq:projection}
$
where $W_x \!\in\! \mathbb{R}^{d_h \times d_x}$ and 
$W_z \!\in\! \mathbb{R}^{d_h \times d_e}$ are learnable matrices.

\subsubsection{Graph Encoding} 
The relational and temporal dependencies in $G_t$ are aggregated using a multi-layer GNN through message passing:
\begin{equation}
\mathbf{h}_i^{(\ell+1)} =
\sigma \!\left(
\sum_{\tau \in \mathbb{T}}
\sum_{j \in \Gamma_t^{(\tau)}(i)}
\frac{1}{c_{i,\tau}}\,
W_\tau^{(\ell)}
\big[\mathbf{h}_j^{(\ell)} \Vert \tilde{\mathbf{z}}_{ij}\big]
\right),
\label{eq:gnn_update}
\end{equation}
where $\mathbb{T}$ is the set of relation types, 
$\Gamma_t^{(\tau)}(i)$ denotes neighbors of node~$i$ under relation~$\tau$, 
$c_{i,\tau}$ is a normalization factor, and $\sigma(\cdot)$ is a nonlinear activation.  
After $L$ layers, a permutation-invariant attention readout yields the final graph embedding:
\begin{equation}
g_t = 
\text{READOUT}
\!\left(
\{\mathbf{h}_i^{(L)} \mid v_i \in V_t\}
\right),
\quad
g_t \in \mathbb{R}^{d_g}.
\label{eq:readout}
\end{equation}

The resulting $g_t$ compactly encodes both intra-host and inter-host dependencies, serving as the temporal input sequence to the LSTM-based Stage Estimator.

\begin{table}
\centering
\caption{APT stages derived from MITRE ATT\&CK framework \cite{mitre_attack}.}
\label{tab:mitre_stages}
\begin{tabular}{|c|l|l|}
\hline
\textbf{Stage ID} & \textbf{Stage Name} & \textbf{Representative Techniques} \\
\hline
\textit{k}=1 & Reconnaissance & Network scanning \\
\hline
\textit{k}=2 & Initial Compromise & PowerShell payloads \\
\hline
\textit{k}=3 & Privilege Escalation & Registry autoruns \\
\hline
\textit{k}=4 & Lateral Movement & PsExec, WMI \\
\hline
\textit{k}=5 & Command and Control & Beaconing \\
\hline
\textit{k}=6 & Exfiltration & Data compression \\
\hline
\end{tabular}
\end{table}

\subsection{LSTM-Based Stage Estimator}
\label{sec:stage_estimator}

Next, the Stage Estimator identifies the current phase of an APT campaign by analyzing temporal patterns within fused provenance graph embeddings. Guided by the MITRE ATT\&CK Enterprise framework~\cite{mitre_attack}, the attack lifecycle is abstracted into six operational stages, as summarized in Table~\ref{tab:mitre_stages}. Each stage reflects distinctive behavioral and structural signatures, facilitating interpretable and data-efficient classification. A benign class ($k=0$) is also defined to represent normal system operation, completing the stage taxonomy.

Because APTs evolve sequentially over time, we employ a Long Short-Term Memory (LSTM) network to model temporal dependencies in the sequence of graph embeddings 
\(\{ g_{1}, g_{2}, \ldots, g_t \}\), where each \(g_t \in \mathbb{R}^{d_g}\) encodes a fused provenance graph snapshot at time~$t$. 
At each step, the LSTM updates its hidden and cell states as
\begin{equation}
(h_t, c_t) = f_{\text{LSTM}}(g_t, h_{t-1}, c_{t-1}),
\end{equation}
where $h_t$ captures both short- and long-term context across time windows, and $c_t$ represents the internal memory.  

The output hidden state $h_t$ is passed through a softmax classifier to estimate the probability distribution over stages:
\begin{equation}
p_t = \mathrm{softmax}\!\left(W_{\text{stage}} h_t + B_{\text{stage}}\right),
\label{eq:stage_probability}
\end{equation}
where \(p_t^{(k)}\) denotes the likelihood of stage \(k\), \(W_{\text{stage}}\) and \(B_{\text{stage}}\) are learnable parameters.  

The model thus provides a continuous estimation of the attacker’s current phase based on historical system-network behaviors encoded in the fused provenance graphs. 

\subsection{Attack Stage Mapping}
Finally, the Attack Stage Mapping module converts the Stage Estimator’s probabilistic outputs into interpretable attack-phase insights. Given the stage probability vector $p_t$, it determines the most likely phase $\hat{k}_t = \arg\max_k p_t^{(k)}$ and tracks its temporal evolution to reveal transitions across the APT lifecycle. This time-aligned mapping aids analysts in understanding attack progression and can export stage estimates as structured alerts to external defense components, such as response orchestration or threat-hunting systems, thus bridging stage inference with actionable defense.

\section{Performance Evaluation}\label{PerformanceEvaluation}

\subsection{Datasets}
The StageFinder framework is evaluated using two complementary datasets from the DARPA cyber defense programs, which together provide both labeled and large-scale unlabeled telemetry for APT stage inference.

\paragraph{DARPA Transparent Computing (TC)}
The TC dataset \cite{DARPA_TC_Dataset} contains multi-week red/blue-team engagements on instrumented enterprise networks, recording fine-grained system provenance from Linux, FreeBSD, and Windows hosts. Subsets such as CADETS, THEIA, TRACE, and FiveDirections capture process creation, file I/O, IPC, and network events, totaling 200--600 million records across 10--50 hosts per engagement. Each campaign includes ground-truth manifests aligned with MITRE ATT\&CK stages, enabling supervised labeling of provenance graph snapshots for training and evaluation of the Stage Estimator.

\paragraph{DARPA Operationally Transparent Cyber (OpTC)}
The OpTC dataset \cite{darpa_optc_dataset} extends TC by integrating large-scale host and network telemetry across hundreds of systems, containing roughly 8.7\,billion host events and 0.53\,billion Zeek flow logs from multi-day red-team operations. Although OpTC lacks explicit APT stage annotations, it provides correlated, high-volume sequences of host–network interactions suitable for self-supervised pretraining of the Stage Estimator.

\subsection{LSTM-based Stage Estimator Training Strategy}
The Stage Estimator is trained in two sequential phases to exploit both unlabeled and labeled telemetry. First, self-supervised pretraining is performed on the large-scale DARPA~OpTC dataset to learn generic temporal dependencies between host and network activities. Second, the pretrained model is fine-tuned on the DARPA~Transparent Computing (TC) dataset for stage-specific discrimination using ground-truth red-team annotations.

\paragraph{Self-Supervised Pretraining on OpTC}
For each 300\,s window, a fused provenance graph $G_t$ is encoded by the GNN into a vector $g_t=f_{\text{GNN}}(G_t)\!\in\!\mathbb{R}^{d_g}$, forming sequences $\{g_{t-L+1},\ldots,g_t\}$ ($L{=}20$). 
A two-layer LSTM ($H{=}128$, dropout~0.3) is trained with two objectives: next-step prediction
$
\mathcal{L}_{\text{pred}}=\tfrac{1}{T-1}\sum_{t}\!\|g_{t+1}-\hat{g}_{t+1}\|_2^2,\qquad
\hat{g}_{t+1}=W_o h_t+b_o,
$
and a temporal contrastive loss
$
\mathcal{L}_{\text{ctr}}=-\!\tfrac{1}{T-1}\!\sum_{t}
\log\!\frac{\exp(\mathrm{sim}(h_t,g_{t+1})/\tau)}
{\sum_{g^-\!\in\!\mathbb{N}}\exp(\mathrm{sim}(h_t,g^-)/\tau)},
$
with $\tau{=}0.2$ and $|\mathbb{N}|{=}256$. 
The joint objective is $\mathcal{L}_{\text{ssl}}=\lambda_{\text{pred}}\mathcal{L}_{\text{pred}}+\lambda_{\text{ctr}}\mathcal{L}_{\text{ctr}}$.
Training uses Adam ($\text{lr}{=}10^{-3}$, weight decay $10^{-5}$, batch size~64, 20 epochs, gradient clipping $\|\nabla\|_2\!\le5$).

\paragraph{Supervised Fine-Tuning on TC}
The pretrained LSTM is adapted to labeled TC engagements segmented into 300\,s windows with APT stage labels $y_t\!\in\!\{0,\ldots,6\}$. 
A softmax head predicts stage probabilities:
$
p_t=\mathrm{softmax}(W_{\text{stage}}h_t+B_{\text{stage}}), \quad 
p_t^{(k)}=P(\text{stage}=k\mid g_{1:t}),
$
optimized using weighted cross-entropy:
$
\mathcal{L}_{\text{sup}}=-\tfrac{1}{T}\!\sum_{t,k} w_k\, y_t^{(k)}\log(p_t^{(k)}{+}\epsilon),
$
where $w_k$ compensates class imbalance and $\epsilon{=}10^{-8}$. 
Fine-tuning proceeds in two stages: (i) discriminative training (frozen lower LSTM, 10 epochs, $\text{lr}{=}10^{-4}$) and (ii) end-to-end optimization (20 epochs, $\text{lr}{=}5{\times}10^{-5}$). 
Training uses dropout~0.3, early stopping on validation F1, and temporal curriculum learning (sequence length 10→30). 
All experiments are implemented in PyTorch~2.1 with DGL on an NVIDIA RTX~3060 GPU (32\,GB), completing in $\sim$1.5\,hours.

\subsection{Experimental Setup and Baseline Models}
The Stage Estimator in the StageFinder framework was pretrained on the DARPA~OpTC dataset to learn general host--network dynamics and fine-tuned on the labeled DARPA~TC (Engagement 5) dataset. 
Evaluation was conducted on unseen samples from other TC engagements to assess generalization. 
Logs were segmented into 300\,s windows, and a 3-layer GNN produced graph embeddings $g_t$ for the LSTM input.

Performance was measured using Precision, Recall, F1-score, and Accuracy to evaluate detection quality, while AUPR (area under the precision--recall curve) quantified robustness to class imbalance. 
Temporal stability was assessed via Temporal Flip Rate (TFR)—the proportion of adjacent windows with differing stage predictions, where lower values indicate smoother transitions. 
All results are reported as mean~$\pm$~standard deviation across five temporal folds.

\textit{Baseline Methods}: We compare the StageFinder framework with two representative APT-stage classifiers: (\textit{i}) \textit{Cyberian}~\cite{Cyberian} employs an LSTM to model temporal dependencies in host-level events. For fair comparison, we reimplemented it under the same architecture but without fused network alerts or cross-dataset pretraining, training solely on TC dataset. (\textit{ii}) \textit{NetGuardian}~\cite{NetGuardian} uses stage-specific classifiers—e.g., SVMs for Initial Compromise and anomaly detectors for Lateral Movement and C2—trained on TC-derived features. We reproduced this heuristic pipeline to obtain both per-stage and overall F1-scores.

\begin{table}[t]
\centering
\caption{StageFinder performance on DARPA TC (Engagement 5).}
\label{tab:OverallPerformance}
\small
\begin{tabular}{lccc}
\hline
\textbf{Metric} & \textbf{Cyberian} & \textbf{NetGuardian} & \textbf{StageFinder} \\
\hline
Precision & 0.89 $\pm$ 0.02 & 0.92 $\pm$ 0.02 & \textbf{0.96 $\pm$ 0.01} \\
Recall    & 0.90 $\pm$ 0.03 & 0.91 $\pm$ 0.02 & \textbf{0.96 $\pm$ 0.01} \\
F1-Score  & 0.90 $\pm$ 0.02 & 0.92 $\pm$ 0.02 & \textbf{0.96 $\pm$ 0.01} \\
TFR$\downarrow$ & 0.182 $\pm$ 0.015 & 0.160 $\pm$ 0.012 & \textbf{0.125 $\pm$ 0.010} \\
Accuracy  & 0.90 $\pm$ 0.03 & 0.92 $\pm$ 0.02 & \textbf{0.96 $\pm$ 0.01} \\
AUPR      & 0.91 $\pm$ 0.02 & 0.94 $\pm$ 0.02 & \textbf{0.97 $\pm$ 0.01} \\
\hline
\end{tabular}
\end{table}

\subsection{Result Analysis}
Table~\ref{tab:OverallPerformance} summarizes the comparative performance of Cyberian, NetGuardian, and the StageFinder on the DARPA~TC~(E5) dataset. 
The StageFinder consistently outperforms both baselines across all metrics, confirming the benefit of combining fused provenance graph representations with temporal modeling and cross-dataset pretraining.

The StageFinder achieves a macro F1-score of $0.96\!\pm\!0.01$, improving upon Cyberian ($0.90\!\pm\!0.02$) and NetGuardian ($0.92\!\pm\!0.02$) by roughly 6\% and 4\%, respectively. 
Both Precision and Recall reach $0.96$, reflecting enhanced APT detection and fewer false positives. 
Overall Accuracy and AUPR rise to $0.96\!\pm\!0.01$ and $0.97\!\pm\!0.01$, while the Temporal Flip Rate (TFR) drops from $0.182$ (Cyberian) and $0.160$ (NetGuardian) to $0.125$, indicating a 31\% reduction in prediction volatility. 
This smoother evolution of predicted stages suggests that the LSTM effectively models long-term temporal dependencies within the fused graph sequences.

\begin{table}
\centering
\caption{Per-stage F1-scores on DARPA TC (Engagement 5).}
\label{tab:PerStagePerformance}
\small
\setlength{\tabcolsep}{2.8pt} % tighter column spacing for compact fit
\begin{tabular}{c l c c c}
\hline
\textbf{$k$} & \textbf{Stage} & \textbf{Cyberian} & \textbf{NetGuardian} & \textbf{StageFinder} \\
\hline
0 & Normal      & 0.93$\pm$0.02 & 0.95$\pm$0.02 & \textbf{0.97$\pm$0.01} \\
1 & Recon.      & 0.88$\pm$0.03 & 0.91$\pm$0.03 & \textbf{0.94$\pm$0.02} \\
2 & Init. Comp. & 0.89$\pm$0.03 & 0.92$\pm$0.02 & \textbf{0.96$\pm$0.01} \\
3 & Priv. Esc.  & 0.87$\pm$0.04 & 0.90$\pm$0.03 & \textbf{0.95$\pm$0.02} \\
4 & Lat. Move.  & 0.86$\pm$0.05 & 0.91$\pm$0.04 & \textbf{0.96$\pm$0.01} \\
5 & C2          & 0.88$\pm$0.04 & 0.93$\pm$0.03 & \textbf{0.97$\pm$0.01} \\
6 & Exfil.      & 0.85$\pm$0.05 & 0.90$\pm$0.04 & \textbf{0.95$\pm$0.02} \\
\hline
\textbf{Macro Avg.} & -- & 0.88$\pm$0.03 & 0.92$\pm$0.03 & \textbf{0.96$\pm$0.01} \\
\hline
\end{tabular}
\end{table}

Per-stage results in Table~\ref{tab:PerStagePerformance} further demonstrate consistent gains across all APT phases ($k{=}0$--6). 
In early stages (Reconnaissance and Initial~Compromise), F1-scores increase from $0.88$--$0.92$ (baselines) to $0.94$--$0.96$ with StageFinder, aided by early fusion of host and network alerts. 
For intermediate phases (Privilege~Escalation, Lateral~Movement), scores improve from $0.86$--$0.91$ to $0.95$--$0.96$, reflecting stronger capture of inter-host causal dependencies. 
During Command-and-Control and Exfiltration, the estimator attains $0.97$ and $0.95$, surpassing Cyberian ($0.88$/$0.85$) and NetGuardian ($0.93$/$0.90$). 
The macro F1-score increases from $0.88$ and $0.92$ to $0.96$, corresponding to relative gains of 9\% and 4\%. 
These results confirm that the proposed framework delivers more accurate and temporally stable APT stage estimation across the full intrusion lifecycle.

\begin{figure}
\centering
\includegraphics[width=0.5\textwidth]{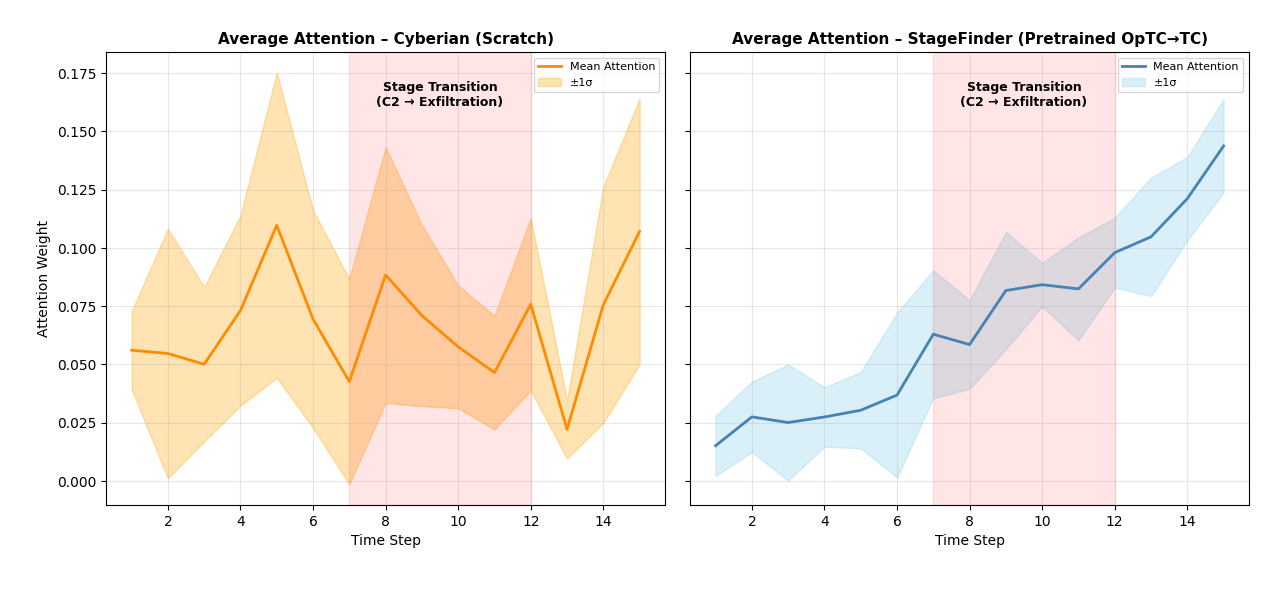}
\caption{Temporal attention comparison between Cyberian and StageFinder LSTM models.}
\label{fig:TemporalAttentionHeatmaps}
\end{figure}

To examine temporal learning behavior, Figure~\ref{fig:TemporalAttentionHeatmaps} depicts the attention distributions of the LSTM models from Cyberian and StageFinder over time. The Cyberian model shows diffuse and irregular attention peaks, suggesting limited ability to consistently focus on semantically relevant temporal segments due to the absence of causal and contextual graph information. In contrast, the StageFinder exhibits concentrated and stable attention patterns aligned with mid-to-late time windows corresponding to the \textit{Command-and-Control} and \textit{Exfiltration} stages. These focused distributions indicate enhanced temporal coherence and clearer recognition of critical APT transitions, confirming the model’s capacity to capture long-range dependencies within fused provenance graph sequences.

\section{Conclusion}\label{Conclusion}
This work presented \textit{StageFinder}, a temporal-graph learning framework for accurate inference of APT attack stages from fused host and network provenance data. 
By integrating early fusion at the graph-construction stage with a GNN-based encoder and an LSTM temporal estimator, StageFinder captures both structural causality and temporal evolution across multi-host environments. 
Evaluations on DARPA~TC and OpTC datasets demonstrate significant improvements in detection accuracy, temporal stability, and stage-level interpretability over state-of-the-art baselines. 
The framework’s modular design allows its outputs to be directly consumed by higher-level defense systems for automated response. 
In our future work, we will extend the framework toward joint stage estimation and adaptive defense policy learning.
 
\section*{Acknowledgment}
This work has been performed in the framework of the SUSTAINET-Advance project, funded by the German BMFTR (ID:16KIS2280).

\bibliographystyle{ieeetr}
\bibliography{References.bib}

\end{document}